%% file: berryconnections.tex
\documentclass{ws-ijmpa}
\usepackage[super,compress]{cite}
\usepackage{hyperref}
\hypersetup{colorlinks,urlcolor=black,citecolor=black,linkcolor=black,filecolor=black}
\usepackage{breakurl}
\usepackage{braket}
\usepackage{mathrsfs}
\usepackage{comment}
\usepackage{tikz}

\input{preamble}

\begin{document}

\thispagestyle{plain}

\def\bib{B\kern-.05em{I}\kern-.025em{B}\kern-.08em}
\def\btex{B\kern-.05em{I}\kern-.025em{B}\kern-.08em\TeX}

\markboth{GLSM@30}{Berry connections, spectral curves and difference equations}

\title{On Spectral Data for $(2,2)$ Berry Connections, Difference Equations \& Equivariant Quantum Cohomology}

\author{Andrea E. V. Ferrari }

\address{School of Mathematics, University of Edinburgh \& DESY\\
Edinburgh, EH9 3FD, United Kingdom \& Notkestraße 85, 22607 Hamburg
\\
andrea.e.v.ferrari@gmail.com}

\author{Daniel Zhang}

\address{St John’s College, University of Oxford, St Giles'\\
Oxford, OX1 3JP, United Kingdom\\
daniel.zhang@sjc.ox.ac.uk}

\maketitle


\begin{abstract}

We study supersymmetric Berry connections of 2d $\mathcal{N}=(2,2)$ gauged linear sigma models (GLSMs) quantized on a circle, which are periodic monopoles, with the aim to provide a fruitful physical arena for recent mathematical constructions related to the latter. These are difference modules encoding monopole solutions via a Hitchin--Kobayashi correspondence established by Mochizuki. We demonstrate how the difference modules arise naturally by studying the ground states as the cohomology of a one--parameter family of supercharges. In particular, we show how they are related to one kind of monopole spectral data, a quantization of the Cherkis--Kapustin spectral curve, and relate them to the physics of the GLSM. By considering states generated by D--branes and leveraging the difference modules, we derive novel difference equations for brane amplitudes. We then show that in the conformal limit, these degenerate into novel difference equations for hemisphere partition functions, which are exactly calculable. When the GLSM flows to a nonlinear sigma model with K\"ahler target $X$, we show that the difference modules are related to the equivariant quantum cohomology of $X$. 

\end{abstract}


\section{Introduction}

Supersymmetric gauge theories in low dimensions have been an inexhaustible source of deep mathematical constructions and problems. This is undoubtedly the case for 2d $\mathcal{N}=(2,2)$ GLSMs, which originated the study of mirror symmetry. This contribution to the proceedings summarizes aspects of the article \cite{ad} by the two authors, demonstrating that this particular source still has much to give. 

We revisit some phenomena related to the supersymmetric ground states of 2d $(2,2)$ GLSMs quantized on a circle, either in a cylindrical or cigar geometry. We consider theories with an abelian flavor symmetry for which a corresponding (generic) twisted mass deformation results in massive topologically trivial vacua. The starting, fundamental observation is that moduli spaces of solutions to supersymmetric Berry connections over a twisted mass deformation and associated holonomy for an abelian flavor symmetry correspond to moduli spaces of periodic monopoles. This allows us to relate supersymmetric ground states in the cohomology of a one--parameter family of supercharges to mathematical constructions that have recently received significant attention, namely difference modules representing monopole solutions due to Mochizuki \cite{mochizuki2017periodic, mochizuki2019doubly}. These can be thought of as encoding a kind of spectral data for the monopole. As a result of this relation, we derive novel difference equations satisfied by brane amplitudes and hemisphere or vortex partition functions. We demonstrate how these difference equations can be thought of as a quantization of the Cherkis--Kapustin spectral curve\cite{Cherkis:2000cj, Cherkis:2000ft} for the monopole. Moreover, in the case of a GLSM that flows to a NLSM with K\"ahler target $X$, we relate these modules to the equivariant quantum cohomology of $X$. To the best of our knowledge, this connection has also not appeared so far in the literature.

In the longer version of this article \cite{ad} we present a generalization to higher rank symmetries \textit{i.e.} for generalized monopoles on $(\bR^2 \times S^1)^n$, $n \geq 1$. This should correspond to an extension of the work of Mochizuki. We also discuss alternative spectral data, which is the analog of the Hitchin\cite{hitchin1982monopoles} spectral curve for monopoles in $\bR^3$. We relate this to the physics of the GLSM and demonstrate how this data is instead related to the equivariant K--theory of $X$. The existence of these alternative descriptions is a physical incarnation of a Riemann--Hilbert correspondence in the style of Kontsevich--Soibelman.\cite{KontsevichSoibelman}

\section{Set--up}\label{sec:set-up}

The set--up of this contribution to the proceedings is 2d $(2,2)$ gauged linear sigma models with a flavor symmetry $T=U(1)$, quantized on the Euclidean cylinder $\mathbb{R}\times S^1$. We turn on a complex twisted mass $w=w_1 + i w_2$ for $T$, and a corresponding flat connection around the $S^1$ with (periodic) holonomy $t = t+L$. Thus, as a real manifold, we consider a space of deformations parameterized by $(t,w)$
\begin{equation}
    M := S^1 \times \mathbb{R}^2.
\end{equation}

The 2d $\mathcal{N}=(2,2)$ algebra contains a family of 1d $\mathcal{N}=(2,2)$ supersymmetric quantum mechanics along $\mathbb{R}$:
\begin{equation}\label{eq:modified_algebra}
\begin{split}
\{Q_{\lambda}, Q_{\lambda}^{\dagger}\} &= \{\bar{Q}_\lambda , \bar{Q}_\lambda^\dagger \} = 2H, \\
\{Q_{\lambda},\bar{Q}_{\lambda}\} =  \tilde{Z}_{t_0}&, \quad Q_{\lambda}^2 = \tilde{Z}_{\beta_0}, \quad \bar{Q}_{\lambda}^2 =  -\tilde{Z}^*_{\beta_0},
\end{split}
\end{equation}
where $\lambda$ is a complex (twistor) parameter (\emph{c.f.} appendix B of~\cite{Gaiotto:2016hvd}) and
\begin{equation}\label{eq:lambda_supercharges}
\begin{split}
Q_{\lambda} = \frac{1}{\sqrt{1+|\lambda|^2}}(Q_A +\lambda \bar{Q}_A), \qquad \bar{Q}_{\lambda} = \frac{1}{\sqrt{1+|\lambda|^2}}(\bar{Q}_A - \bar{\lambda } Q_A), \\
Q_{\lambda}^{\dagger} = \frac{1}{\sqrt{1+|\lambda|^2}}(Q_A^{\dagger} +\bar{\lambda} \bar{Q}_A^{\dagger}), \qquad \bar{Q}_{\lambda}^{\dagger} = \frac{1}{\sqrt{1+|\lambda|^2}}(\bar{Q}_A^{\dagger} - \lambda  Q_A^{\dagger}).
\end{split}
\end{equation}
Here $Q_A=Q_{\lambda}|_{\lambda = 0}$ is the supercharge whose cohomology defines the A--twist, and $Q_A^\dagger$ its hermitian conjugate. On gauge--invariant states and operators the central charges are given by
\begin{equation}
    \tilde{Z}_{t_0} = \frac{1-|\lambda|^2}{1+|\lambda|^2} (2i \partial_2) + 2 t_0\cdot J_T, \quad \tilde{Z}_{\beta_0} =  \frac{2i \lambda \partial_2}{1+|\lambda|^2} -i\beta_0 \cdot J_T
\end{equation}
with $\partial_2$ the momentum along $S^1$, $J_T$ a generator for $T$ and
\begin{equation}\label{eq:par-dep}
    \begin{split}
       t_0 =  \frac{1-|\lambda|^2}{1+|\lambda|^2} t + \frac{2}{1+|\lambda|^2} \Im (\lambda \bar w), \quad 
       \beta_0 =\frac{1}{1+|\lambda|^2}(w+\lambda^2\bar{w} + 2i\lambda t).
    \end{split}
\end{equation}

We will be interested in the space of supersymmetric ground states for these SQMs, defined as follows. First, we require the vanishing $\tilde{Z}_{t_0} = \tilde{Z}_{\beta_0} = 0$ of central charges on the space of ground states. This generically implies that ground states are uncharged under $J_T$, and have no KK modes along the circle. Then, on the states satisfying this condition we can further impose $H = 0$. Consider fixed values of deformation parameters $(t,w)$ and denote the vector space of supersymmetric ground states by $E$ as a subspace of the states of the theory $S$. Then whenever the system is gapped we have a cohomological description of the ground states
\begin{equation}
    E \cong H^\bullet ( S|_{\tilde{Z}_{t_0} = \tilde{Z}_{\beta_0}=0}, Q_\lambda).
\end{equation}

\subsection{Mini--complex structures on the space of deformations}
\label{subsec:def-par}

The family of SQMs is related to a one--parameter family of mini--complex structures on the parameter space $M=S^1\times \mathbb{R}^2$. These were defined by Mochizuki~\cite{mochizuki2017periodic}, and a precise definition is beyond the scope of this contribution (see \textit{e.g.}\cite{ad}). However, it suffices to say that a mini--complex structure ensures the three--manifold has a collection of charts of the form $\mathbb{R} \times \mathbb{C}$, so that there is a meaningful notion of functions that are locally constant along $\mathbb{R}$ and holomorphic along $\mathbb{C}$.

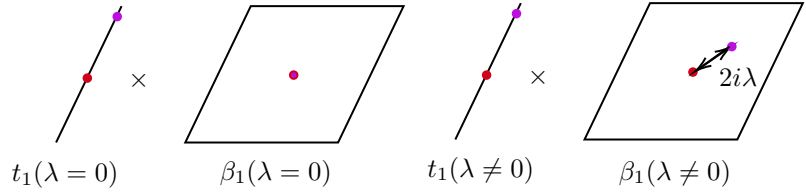
\begin{figure}
    \centering
    
\tikzset{every picture/.style={line width=0.75pt}} 
\begin{tikzpicture}[x=0.75pt,y=0.75pt,yscale=-0.75,xscale=0.75]

\draw    (489,59.5) -- (511.35,44.13) ;
\draw [shift={(513,43)}, rotate = 145.49] [color={rgb, 255:red, 0; green, 0; blue, 0 }  ][line width=0.75]    (10.93,-3.29) .. controls (6.95,-1.4) and (3.31,-0.3) .. (0,0) .. controls (3.31,0.3) and (6.95,1.4) .. (10.93,3.29)   ;
\draw    (372.4,14) -- (328.6,105) ;
\draw   (458.8,14) -- (561,14) -- (517.2,105) -- (415,105) -- cycle ;
\draw  [color={rgb, 255:red, 208; green, 2; blue, 27 }  ,draw opacity=1 ][fill={rgb, 255:red, 208; green, 2; blue, 27 }  ,fill opacity=1 ] (485,60) .. controls (485,58.62) and (486.12,57.5) .. (487.5,57.5) .. controls (488.88,57.5) and (490,58.62) .. (490,60) .. controls (490,61.38) and (488.88,62.5) .. (487.5,62.5) .. controls (486.12,62.5) and (485,61.38) .. (485,60) -- cycle ;
\draw  [color={rgb, 255:red, 208; green, 2; blue, 27 }  ,draw opacity=1 ][fill={rgb, 255:red, 208; green, 2; blue, 27 }  ,fill opacity=1 ] (347,62) .. controls (347,60.62) and (348.12,59.5) .. (349.5,59.5) .. controls (350.88,59.5) and (352,60.62) .. (352,62) .. controls (352,63.38) and (350.88,64.5) .. (349.5,64.5) .. controls (348.12,64.5) and (347,63.38) .. (347,62) -- cycle ;
\draw  [color={rgb, 255:red, 189; green, 16; blue, 224 }  ,draw opacity=1 ][fill={rgb, 255:red, 189; green, 16; blue, 224 }  ,fill opacity=1 ] (511,43) .. controls (511,41.62) and (512.12,40.5) .. (513.5,40.5) .. controls (514.88,40.5) and (516,41.62) .. (516,43) .. controls (516,44.38) and (514.88,45.5) .. (513.5,45.5) .. controls (512.12,45.5) and (511,44.38) .. (511,43) -- cycle ;
\draw  [color={rgb, 255:red, 189; green, 16; blue, 224 }  ,draw opacity=1 ][fill={rgb, 255:red, 189; green, 16; blue, 224 }  ,fill opacity=1 ] (367,21) .. controls (367,19.62) and (368.12,18.5) .. (369.5,18.5) .. controls (370.88,18.5) and (372,19.62) .. (372,21) .. controls (372,22.38) and (370.88,23.5) .. (369.5,23.5) .. controls (368.12,23.5) and (367,22.38) .. (367,21) -- cycle ;
\draw    (105.4,16) -- (61.6,107) ;
\draw   (191.8,16) -- (294,16) -- (250.2,107) -- (148,107) -- cycle ;
\draw  [color={rgb, 255:red, 208; green, 2; blue, 27 }  ,draw opacity=1 ][fill={rgb, 255:red, 189; green, 16; blue, 224 }  ,fill opacity=1 ] (218,62) .. controls (218,60.62) and (219.12,59.5) .. (220.5,59.5) .. controls (221.88,59.5) and (223,60.62) .. (223,62) .. controls (223,63.38) and (221.88,64.5) .. (220.5,64.5) .. controls (219.12,64.5) and (218,63.38) .. (218,62) -- cycle ;
\draw  [color={rgb, 255:red, 208; green, 2; blue, 27 }  ,draw opacity=1 ][fill={rgb, 255:red, 208; green, 2; blue, 27 }  ,fill opacity=1 ] (80,64) .. controls (80,62.62) and (81.12,61.5) .. (82.5,61.5) .. controls (83.88,61.5) and (85,62.62) .. (85,64) .. controls (85,65.38) and (83.88,66.5) .. (82.5,66.5) .. controls (81.12,66.5) and (80,65.38) .. (80,64) -- cycle ;
\draw  [color={rgb, 255:red, 189; green, 16; blue, 224 }  ,draw opacity=1 ][fill={rgb, 255:red, 189; green, 16; blue, 224 }  ,fill opacity=1 ] (100,23) .. controls (100,21.62) and (101.12,20.5) .. (102.5,20.5) .. controls (103.88,20.5) and (105,21.62) .. (105,23) .. controls (105,24.38) and (103.88,25.5) .. (102.5,25.5) .. controls (101.12,25.5) and (100,24.38) .. (100,23) -- cycle ;
\draw    (512,43) -- (490.63,58.33) ;
\draw [shift={(489,59.5)}, rotate = 324.34] [color={rgb, 255:red, 0; green, 0; blue, 0 }  ][line width=0.75]    (10.93,-3.29) .. controls (6.95,-1.4) and (3.31,-0.3) .. (0,0) .. controls (3.31,0.3) and (6.95,1.4) .. (10.93,3.29)   ;

\draw (375,55.4) node [anchor=north west][inner sep=0.75pt]    {$\times $};
\draw (308,114.4) node [anchor=north west][inner sep=0.75pt]    {$t_{1}( \lambda \neq 0)$};
\draw (436,116.4) node [anchor=north west][inner sep=0.75pt]    {$\beta _{1}( \lambda \neq 0)$};
\draw (108,57.4) node [anchor=north west][inner sep=0.75pt]    {$\times $};
\draw (30,117.4) node [anchor=north west][inner sep=0.75pt]    {$t_{1}( \lambda =0)$};
\draw (169,116.4) node [anchor=north west][inner sep=0.75pt]    {$\beta _{1}( \lambda =0)$};
\draw (503,54.65) node [anchor=north west][inner sep=0.75pt]    {$2i\lambda $};

\end{tikzpicture}
    \caption{The mini--holomorphic coordinates $(t_1,\beta_1)$ at different $\lambda$. The purple and red points are identified in the underlying smooth manifold $M \cong S^1\times \mathbb{R}^2$. In the product case ($\lambda=0$, left), moving along the real coordinate brings one back to the same point in $M$. In the non--product case ($\lambda \neq 0$, right), an additional shift by $2i\lambda$ is necessary. }
    \label{fig:intro-mini-coord}
\end{figure}

In the case at hand, the mini--complex structures come from a lift $t \in \mathbb{R}$ (abusing notation), so that the parameter space is diffeomorphic to $\mathbb{R}^3$. Lifts of $(t_0,\beta_0)$ deliver obvious mini--complex structures in that we can view $\mathbb{R}^3 \cong \mathbb{R}_{t_0}\times \mathbb{C}_{\beta_0}$. $M$ can then be recovered as the quotient of this mini--complex manifold by a $\mathbb{Z}$ action that makes $t$ periodic. There are two qualitatively distinct cases, as depicted in Figure~\ref{fig:intro-mini-coord}:
\begin{itemize}
    \item The first case, also known as the product case, is characterized by $\lambda = 0$ so that $(t,w)\sim (t+L,w)$. Thus, $M \cong S^1 \times \mathbb{C}$ as a mini--complex manifold.
    \item The second case, also known as the non--product case, is characterized by $\lambda \neq 0 $ so that the $(t_0, \beta_0) \sim  (t_0, \beta_0) + \frac{L}{1+|\lambda|^2}(1-|\lambda|^2, 2i\lambda)$.
\end{itemize}

In\cite{mochizuki2017periodic} Mochizuki also introduces a second set of mini--complex coordinates that are closely related to $(t_0,\beta_0)$. For later convenience, we report them here:
\begin{equation}\label{eq:mochizukisecondcoords}
(t_1,\beta_1) = (t_0 + \text{Im}(\bar{\lambda}\beta_0),  (1+|\lambda|^2)\beta_0) = (t+ \text{Im}(\lambda\bar{w}), w + 2 i \lambda t + \lambda^2 \bar{w}).
\end{equation}
This can be thought as another coordinate system for the same mini--complex structure parameterized by $\lambda$, obeying $(t_1,\beta_1) \sim (t_1,\beta_1) + L(1,2i\lambda)$.

\section{Berry Connections \& Asymptotics}

In this section, we introduce the vector bundle of supersymmetric ground states $E \rightarrow M$. We work with coordinates $x=(t,w) \in M$. We review features of $tt^*$ geometry for a twisted mass deformation. For simplicity, we will usually take the quantization circle to have length $L=1$, re--introducing it where necessary. 

\subsection{\texorpdfstring{$tt^*$ }{}geometry}

For theories with $N$ vacua, there is a $U(N)$ Berry connection on $E$. This is the Berry connection for a twisted mass deformation and accompanying holonomy.\cite{Cecotti:2013mba} The connection itself is defined in the usual way, where if $\ket{\alpha(x)}$ denotes an orthonormal basis of ground states at parameter value $x$:
\begin{equation}
    \partial_x \ket{\alpha(x)} = (A_x)_{\alpha}{}^{\beta} \ket{\beta(x)}.
\end{equation}

In the present case of a rank one flavor symmetry, so that $M = S^1 \times \mathbb{R}^2$, the $tt^*$ equations may be equivalently written as:
\begin{equation}\label{bogomolny}
\begin{aligned}
[\bar{D}_{\bar{w}},D_t - i\phi ]   = & \,0 = [D_{w},D_t + i\phi ], \\
 2 [D_w,  \bar{D}_{\bar{w}}] &= i [D_t,\phi],
\end{aligned}
\end{equation}
where
\vspace*{-2mm}
\begin{equation}
    D_w= \partial_{w}-A_{w},\quad \bar{D}_{\bar{w}} = \bar{\partial}_{\bar{w}}-\bar{A}_{\bar{w}},
\end{equation}
and $\phi$ is an anti--Hermitian adjoint Higgs field~\cite{Cecotti:2013mba}. These are simply the Bogomolny equations on $\mathbb{R}^2\times S^1$
\begin{equation}\label{eq:bogomolny}
	F(D) = \star D \phi
\end{equation}
where $F(D)$ is the curvature. That $D$ satisfies \eqref{eq:bogomolny} may also derived as a consequence of a theory reducing to a 1d $\cN=(2,2)$ quantum mechanics. \cite{Sonner:2008fi} 

In summary, the ground state structure over $M$ can be packaged into a tuple $(E,h,D ,\phi)$ consisting of a vector bundle $E$ of ground states, with Hermitian metric $h$ determined by the inner product, a connection $D$ that is unitary with respect to $h$ and an anti--Hermitian endomorphism $\phi$ of $(E,h)$. The tuple satisfies the Bogomolny equation, and may be regarded as a monopole on $M \cong S^1 \times \mathbb{R}^2$.

\subsection{Asymptotics} 

In this work, we will be concerned only with periodic monopoles of generalized Cherkis--Kapustin (GCK) type \cite{Cherkis:2000cj, Cherkis:2000ft}, as coined in\cite{mochizuki2017periodic}. It is shown therein that they are in one--to--one correspondence with certain difference modules, which play a central role in our paper. 

A monopole is of GCK--type if it has Dirac--type singularities at a discrete finite subset $Z\subset M$ and satisfies the following conditions
\begin{equation}\label{eq:GCK_condition}
| \phi_x| = O(\log (d(x,x_0))),~|F(D)_x|\rightarrow 0
\end{equation}
for some reference point $x_0$ as $x$ goes to infinite distance.

Such conditions are satisfied for the basic periodic Dirac monopole \cite{Cherkis:2000cj} of charge $k$, for which the Higgs field $\phi$ satisfies the Laplace equation:
\begin{gather}\label{eq:dirac_monopole_asymptotics}
     i \phi =    c_1 +  \frac{ \gamma k }{2} + \frac{ k }{2} \sum_{n \in \bZ}'\left( \frac{1}{\sqrt{|w|^2+(t-n)^2}} - \frac{1}{|n|}\right) \rightarrow c_1 +  k\log \left|\frac{iw}{2}\right|+o(1)\\
     A_t \rightarrow  i c_2 + i k \arg \left(iw/2\right) + o(1),  \quad A_w \rightarrow b/w + o(1/w)\nonumber
\end{gather}
as $|w| \rightarrow \infty$, where $\gamma$ is the Euler constant, and $'$ on the sum means that for $n=0$ the second term in the summand is omitted. Here, $b,c_1$ and $c_2$ are real constants. For $ c_1 = c_2 = 0$, $b = -\frac{1}{4}$ this is the Berry connection for the free $\mathcal{N}=(2,2)$ chiral \cite{Cecotti:2013mba}.

The GCK conditions \eqref{eq:GCK_condition} are satisfied for the Berry connections of the theories we consider in this work. This is because we assume that as $w \rightarrow \infty$ in a generic direction in $\mathbb{C}$, the theory is fully gapped with massive topologically trivial vacua. The theory will fail to be gapped only at a discrete finite subset of points in the parameter space, corresponding to $Z$ above. Thus, asymptotically in $w$, the $U(N)$ vector bundle $E$ splits
into a direct sum of $U(1)$ bundles $\bigoplus_{\alpha} E_{\alpha}$, each corresponding to a decoupled sector for the effective theory of massive chiral multiplets parametrising perturbations around a massive vacuum. It follows that the solution is asymptotically gauge--equivalent to an abelian solution $(A^{\alpha}, \phi^{\alpha})$ of Dirac monopole solutions with particular moduli determined by the theory. 
 
The asymptotic value of $A_{t}^{\alpha} + i \phi^{\alpha}$ is determined by the dependence of the twisted central charge $\tilde{\mathcal{Z}}_{\alpha}$ in the vacuum $\alpha$ on $w$. For a GLSM, this can be evaluated as the value of the effective (twisted) superpotential $W_{\text{eff}}$ (appearing in the low--energy theory on the Coulomb branch) in the vacuum $\alpha$~\cite{Witten:1993yc}
\begin{equation}
    A_{t}^{\alpha} + i \phi^{\alpha}\sim -2 i \partial_{w} W_{\text{eff}}^{(\alpha)} \quad \text{as} \quad { |w| \rightarrow \infty.}
\end{equation}
This was demonstrated for LG theories,\cite{Cecotti:2013mba} where $W_{\text{eff}}^{(\alpha)}$ can be evaluated as the superpotential at the critical point $\alpha$. The statement for GLSMs is simply the mirror dual of this. This can also be demonstrated for $\mathcal{N}=(2,2)$ quantum mechanics~\cite{Sonner:2008fi} where the twisted central charge is given by the moment map for the corresponding flavor symmetry. In 2d, these are quantum corrected to $W_{\text{eff}}$ via integrating out massive chiral multiplets.

\subsubsection{Example: supersymmetric QED \& \texorpdfstring{$\mathbb{CP}^1$ $\sigma$}{}-model}\label{subsubsec:sqed2_asymptotics}

We take as a running example throughout this work supersymmetric QED with two chiral multiplets, which engineers the $\mathbb{CP}^1$ $\sigma$-model in the IR. This is a $G = U(1)$ GLSM with two chiral multiplets $\Phi_1$, $\Phi_2$ of charges $(+1,\pm1)$ under $G\times T$. We turn on a mass $m = i w /2$ for $T$, and study the Berry connection over $m$ and the associated holonomy $t$, which is a smooth $SU(2)$ monopole solution. \cite{Cecotti:2013mba}

The effective twisted superpotential of the theory is given by
\begin{equation}\label{eq:w_eff_sqed2}
W_{\text{eff}} = -2 \pi i \tau(\mu)  \sigma + (\sigma + m)\Big(  \log  \Big(\frac{\sigma + m}{\mu}\Big)  -1 \Big) +(\sigma - m)\Big(  \log  \Big(\frac{\sigma - m}{\mu}\Big)  -1 \Big).
\end{equation}
Here $\sigma$ is the complex scalar in the $U(1)$ vector multiplet, $\tau(\mu)$ the renormalized complex FI parameter $\tau(\mu)= \tau_0 + \frac{2}{2\pi i }\log(\Lambda_0/\mu)$, $\Lambda_0$ some fixed UV energy scale, and $\mu$ the RG scale. $\tau_0 = \frac{\theta}{2\pi} + i r_0$ is the bare parameter, where $\theta$ is the instanton angle and $r$ the bare real FI. For convenience, we define the RG invariant combination
\begin{equation}
    q=   \Lambda_0^{\,2} \,  e^{2 \pi i\tau_0}.
\end{equation}

The vacuum equations are 
\begin{equation}
1 = e^{\frac{\partial W}{\partial \sigma}} = q^{-1} (\sigma+m)(\sigma-m),
\end{equation}
which yield solutions $\sigma = \pm\sqrt{m^2 + q}$ corresponding to the two vacua. Labeling the vacua this way requires a choice of branch cut for the square root, so we will find it instructive to instead label the vacua by $\alpha = 1,2$ where
\begin{equation}
\text{As } q \rightarrow 0: \qquad
\begin{aligned}
    (1): \quad \sigma &=+ \,m + q/(2m)+ O(m^{-2}),\\
    (2): \quad \sigma &=- \,m - q/(2m) + O(m^{-2}).
\end{aligned}
\end{equation}
These correspond to the two fixed points of the the $\mathbb{CP}^1$ sigma model, which to leading order in $q$ are the classical values $\sigma$ take in order for $\Phi_1$ or $\Phi_2$ to acquire a VEV. We therefore have:
\begin{equation}\label{eq:cp1_asymptotics}
\text{As } |m| \rightarrow \infty: \qquad
\begin{aligned}
    \partial_m W_{\text{eff}}^{(1)} &\rightarrow +2\log(2m /\Lambda_0 )-2\pi i \tau_0+O(m^{-2}),\\
    \partial_m W_{\text{eff}}^{(2)}&\rightarrow -2\log(2m /\Lambda_0)+2\pi i \tau_0+O(m^{-2}).\\
\end{aligned}
\end{equation}
The asymptotics are those of two Dirac monopoles with charges $k= \pm2$ and $c_1+ic_2 = \mp 2\pi i (\tau_0 + \frac{2}{2\pi i } \log \Lambda_0)$, where the argument of the log in \eqref{eq:cp1_asymptotics} is now twice $m$. This is consistent as for large $m$ the vacua correspond to the poles of $\mathbb{CP}^1$ and the effective theories in the neighborhood of the vacua are those of chirals of effective masses $\pm 2m$ and charge $\pm 2$ parametrising the tangent spaces $T_{(1)}\mathbb{CP}^1$ and $T_{(2)}\mathbb{CP}^1$.

\section{Spectral Data \& Difference Modules}

One fundamental question for the bundle of supersymmetric ground states is how the supercharges behave with respect to changes in deformation parameters. In particular, $Q_{\lambda}$ is a B--type supercharge with respect to $\beta_1$ and A--type with respect to $t_1$. This means that the supercharges $Q_{\lambda}$ have the following explicit dependencies
\begin{equation}\label{eq:supercharge_dependencies}
\partial_{\bar{\beta_{1}}}Q_{\lambda} = 0,\quad 
\partial_{t_1} Q_{\lambda} - i[\phi,Q_{\lambda}] = 0,
\end{equation}
where $\phi$ is an anti--Hermitian operator (the Higgs field in the Bogomolny equations). The above follow from the dependencies of $Q_A, \bar{Q}_A$ on $(w,t)$, which are simply the above equations evaluated at $\lambda=1$ and $\lambda = \infty$. This is equivalent to saying that for the A--supercharge basis, the twisted mass deformation is of BAA--type.\cite{Gaiotto:2016hvd}

Equation \eqref{eq:supercharge_dependencies} implies that the anti--holomorphic derivative commutes with the supercharge and descends to a holomorphic Berry connection  $\partial_{E,\bar{\beta_1}}$ on ground states
\begin{equation}\label{eq:mochizukioperator1}
    \partial_{E,\bar{\beta}_1}  = \frac{1}{1+|\lambda|^2} \Big( \bar{D}_{\bar{w}} + \frac{i\lambda}{2}(D_{t}+i\phi)\Big).   
\end{equation}
Further, there is a complexified flat connection on the space of ground states
\begin{equation}\label{eq:mochizukioperator2}
    \partial_{E, t_{1}} \coloneqq D_{t_1}  - i\phi =  \frac{1}{1+|\lambda|^2} \left( (1-|\lambda|^2)D_t -2 i \lambda D_w + 2 i \bar{\lambda} \bar{D}_{\bar{w}}\right) - i \phi,
\end{equation}
which also commutes with $Q_{\lambda}$. These operators are thus well--defined on $Q_{\lambda}$-cohomology. Crucially, the $tt^*$ equations also imply commutativity
\begin{equation}\label{eq:commutator}
\left[\partial_{E, t_{1}} , \partial_{E,\bar{\beta}_1} \right] = 0.
\end{equation}
We now demonstrate how these structures on ground states viewed as $Q_{\lambda}$-cohomology classes naturally correspond to $2i\lambda$-difference modules built from $E$. We start with the product case, $\lambda =0$.

\subsection{\texorpdfstring{$\lambda=0$}{}: product case \& Cherkis--Kapustin spectral curve}\label{sec:product_case}

We now explain how we can obtain certain $0$-difference modules from the space of supersymmetric ground states viewed as $Q_A$-cohomology. Given $\lambda=0$, we work with the respective mini--complex coordinates $(t,w)$. More precisely, what we shall obtain is a $0$-difference $\mathbb{C}(w)$-module. This is a pair $(V,F)$ consisting of a  finite dimensional $\mathbb{C}(w)$-module $V$ together with a $\mathbb{C}(w)$-linear automorphism
\begin{equation}
  F : V\rightarrow V.
\end{equation}
Here, $\mathbb{C}(w)$ denotes the field of rational functions of $w$. The construction of these modules constitutes the initial part of the remarkable Hitchin--Kobayashi correspondence established by Mochizuki~\cite{mochizuki2017periodic}, which we can approximately state as follows: there is a bijective correspondence between isomorphism classes of periodic monopoles of GCK--type on $S_L^1\times \mathbb{C}$ and  polystable, parabolic, filtered $0$-difference modules. In this article we only focus on the 0-difference structure.

Consider the differential operators \eqref{eq:mochizukioperator1}, \eqref{eq:mochizukioperator2} at $\lambda=0$
\be
	\partial_{E,t} := D_t - i\phi,  \qquad \partial_{E,\bar{w}} := \bar{D}_{\bar{w}}.
\ee
For each  $0 \leq t^0 \leq L$ we can define a holomorphic vector bundle $\mathcal{E}^{t^0}$ on $\{t^0\}\times \mathbb{C}_w$ 
\begin{equation}
    \cE^{t} := (E|_{\{t\} \times\bC_{w}}, \partial_{E,\bar{w}} ).
\end{equation}
The holomorphic sections $\ket{a(w)}$ of $\cE^{t^0}$ satisfy $\partial_{E,\bar{w}}\ket{a(w)} = 0$. 
To first approximation, the module can be taken to be the space of holomorphic sections at $t=0$
\begin{equation}
    V_{\text{na\"ive}} := H^0 (\mathbb{C}_w,\mathcal{E}^0) \otimes_{\mathbb{C}[w]}\mathbb{C}(w).
\end{equation}
As usual in the context of a Hitchin--Kobayashi correspondence, for the purposes of assigning algebraic data to a solution of the Bogomolny equations we may complexify the gauge group $U(N)$ to $GL(N,\mathbb{C})$ and focus (up to the imposition of a stability condition, which we ignore) on the complex equation
\begin{equation}\label{eq:comm-bog}
	[\partial_{E,t} , \partial_{E,\bar{w}}] = 0.
\end{equation}
This implies that the parallel transport of a section $\ket{a}$ along the $S^1_L$ direction can be performed whilst preserving holomorphicity. In particular, by parallel transporting from $0$ to $L$ we can define the desired $\mathbb{C}(w)$-automorphism of $V_{\text{na\"ive}}$
\begin{equation}
    F: V_{\text{na\"ive}} \rightarrow V_{\text{na\"ive}},
\end{equation}
In other words, the automorphism comes from considering the holonomy of the connection $\partial_{E,t}$ around $S^1_L$. In local coordinates, we may write
\begin{equation}\label{eq:physical-F-def}
   F(w) = \exp{\oint dt \left(A_{t}+i\phi\right)}.
\end{equation}
Notice that $F(w)$ may in general (in the presence of point--like singularities) be a meromorphic function of $w$, and so we get a $\mathbb{C}(w)$-module structure. 

This na\"ive picture can be turned into a rigorous one by considering the behavior at $w\rightarrow \infty$ as well as allowing for meromorphic singularities. \cite{mochizuki2017periodic} Then, $(V,F)$ constitutes the corresponding $0$-difference module.

The associated \textit{spectral curve} $\mathcal{L}$, first considered for $n=1$ by Cherkis and Kapustin~\cite{Cherkis:2000cj, Cherkis:2000ft}, is the Lagrangian submanifold of $(\bC^*)\times \bC$ defined by
\begin{equation}\label{eq:rank_1_spectral_curve}
    \mathcal{L} = \left\{ (p,w) \,|\, \text{det}( p\mathbf{1} - F(w)) = 0 \right\}.
\end{equation}
It is Lagrangian with respect to the holomorphic symplectic form $\frac{dp}{p} \wedge dw$. $\mathcal{L}$ is an $N$-sheeted cover of $\bC_w$, and is equipped with a coherent sheaf $\mathcal{M}$, whose stalks are the eigenspaces of $F$. The pushforward of $\mathcal{M}$ under the projection $\pi: \mathcal{L} \rightarrow \bC_w$ is a rank $N$ holomorphic vector bundle, and coincides with $E_{\{0\}\times \bC_w}$. The corresponding values of $p$ on $\cL$ encode the parallel transport with respect to $D_{t} - i \phi$.

\subsubsection{Physical constructions}\label{sec:physical_examples}

We now describe specifically how one may recover the above structures physically, for a GCK monopole arising as the supersymmetric Berry connection for a GLSM.

Let us consider the states $\ket{a}$ obtained on the boundary $S^1$ of an A--twisted cigar, by inserting an operator $\cO_a$ in $Q_A$-cohomology (an element of the twisted chiral ring). These states will be in $Q_A$-cohomology, and can be projected onto ground states via stretching the topological path integral, implementing a Euclidean time evolution $e^{-\beta H}$ with $\beta \rightarrow \infty$. One can generate a basis for the space of ground states via a basis of the twisted chiral ring in this way, and working with respect to such a basis is often called working in \textit{topological gauge}.\cite{Cecotti:2013mba} In particular, it is a standard result\cite{Cecotti:1991me} of $tt^*$ that in this basis $(\bar{A}_{\bar{w}})_a{}^{b}=0$ and thus
\begin{equation}
    \partial_{E,\bar{w}} \ket{a(w)} = 0.
\end{equation}
Thus, such states can be identified as generating a basis of the module $V$.

The automorphism $F$ also admits a clean interpretation. Recall the origin of $A_t + i \phi$ in the $tt^*$ equations as the chiral ring matrix, describing the action of the $tt^*$-dual operator to $w$ (the operator to which $w$ couples) on the ground states. As $w$ is the complex scalar component of a background vector multiplet for $T$, this is the defect operator inserting a unit of flux for the $T$ gauge field, or alternatively winding the holonomy. The action of $F$ on $V$ corresponds precisely to the action of such defects, which due to topological invariance can be localized to a local operator.

There is another way of seeing this, which further allows an explicit computation of $F(w)$. Consider an effective description of the theory as an abelian theory in the IR after integrating out all the chiral multiplets.\cite{Witten:1993yc} This theory is determined by $W_{\text{eff}}(\sigma, w)$, the effective twisted superpotential, with $\sigma_i$ parametrising the Cartan of the complex scalar in the vector multiplet of the GLSM. In this description, the twisted chiral ring is represented by gauge--invariant polynomials in $\sigma_i$, subject to the ring relations $\exp{\partial_{\sigma_i}W_{\text{eff}}}=1$. From this perspective, the dual operator to $w$ is simply $-2 i \partial_w W_{\text{eff}}$, and from the form of the effective action, see \textit{e.g.} section 7.1.2 of \cite{Closset_2019}, the operator
\vspace*{-3mm}
\begin{equation}\label{eq:flux_operator}
    p = e^{-2 i \frac{ \partial W_{\text{eff}}(\sigma, w)}{ \partial  w }}
\end{equation}
corresponds precisely to the insertion of a unit of flux in the path integral.

To compute $F$ using this description, write $\ket{a} = \cO_a \ket{1}$, where $\ket{1}$ is the state generated by the $A$--twisted cigar path integral with no insertions, and $\cO_a$ is a polynomial in $\sigma$. This notation makes sense because in the twist $\cO_a$ may be brought to act on the boundary. We suppress the $w$-dependence for clarity. The action of $F$ may now be derived by multiplying $\cO_a$ by $p$ in \eqref{eq:flux_operator}. This naively yields an operator rational in $\sigma$, but by consistency must be able to be brought back into the $ \{ \cO_a \}$ basis by identifications using the vacuum equations $\exp{\partial_{\sigma_i}W_{\text{eff}}}=1$. Performing these, we have
\vspace*{-1mm}
\begin{equation}\label{eq:0_difference}
    p\, \cO_a \ket{1} = F_a{}^b \cO_b \ket{1},
\end{equation}
yielding the automorphism $F(w)$ in the basis $\ket{a}$ generated by twisted chiral ring elements $\cO_a$.\footnote{The above arguments can also be made also via Coulomb branch localization, where the chiral ring insertions concretely take the form of polynomial insertions in a contour integral over $\sigma$. } We see an example of this below.

Let us now show how to derive the spectral curve in terms of the physical data, which does not require performing the above substitutions. Note that \eqref{eq:comm-bog} is independent of the radius $L$ of quantization, which simply rescales $D_{t} - i\phi $. Thus, the eigenvalues of $F$ can be computed in the flat space limit $L \rightarrow \infty $. There, outside of codimension-1 loci in $w$ space, the ground states are simply the massive vacua of the theory. 
In this basis, $A_{t}+i\phi$ is given by the VEVs of the aforementioned defect operator for the flavor symmetry $T$ in the massive vacua $\{\alpha\}$, which may in turn be expressed via the low energy effective twisted superpotential:
\begin{equation}\label{eq:0_diff_module}
    A_{t}+i\phi = \text{diag}_{\{\alpha\}}\left( -2 i \partial_{w} W_{\text{eff}}^{(\alpha)} \right) = \text{diag}_{\{\alpha\}}\left( \partial_{m} W_{\text{eff}}^{(\alpha)} \right).
\end{equation}
Here we have introduced the redefined mass $m = i w/2$. Thus in the $L\rightarrow \infty$ limit
\begin{equation}
    F_i(w) = \text{diag}_{\{\alpha\}}\left( e^{\partial_{m} W_{\text{eff}}^{(\alpha)}} \right).
\end{equation}
Therefore, in the case of GLSMs, the spectral curve equations can be written as:
\begin{equation}\label{eq:glsm_spectral_curve}
\begin{aligned}
    e^{\frac{\partial W_{\text{eff}}(\sigma, m)}{\partial_{m} }} &= p,\qquad 
    e^{\frac{\partial W_{\text{eff}}(\sigma, m)}{\partial_{\sigma_i} }} &= 1, \quad\, i=1\ldots r.
\end{aligned}
\end{equation}
Eliminating $\sigma$ from the combined system \eqref{eq:glsm_spectral_curve} yields the variety $\cL$. The spectral curve has been studied for LG models\cite{Cecotti:2013mba} and in a different context for 3d theories \cite{Dimofte:2011ju,Dimofte:2011jd, Bullimore:2014awa}.

\subsubsection{Relation to quantum equivariant cohomology}\label{subsec:reln_quantum_cohom}

The twisted chiral ring is known to reproduce the \textit{quantum equivariant cohomology} $QH^\bullet_T(X)$ of the vacuum manifold $X$. This is a deformation of the cohomology ring via higher degree pseudo--holomorphic curve contributions to correlators \cite{Vafa:1991uz}. It has an alternative description in the IR effective abelian theory\cite{Nekrasov:2009uh} as the ring of Weyl--invariant polynomials in the scalars $\sigma$, subject to the vacuum equations, $e^{\partial_{\sigma_i} W_{\text{eff}}(\sigma, m)} = 1$.  Our analysis therefore shows that the difference module $(V,F)$ can be interpreted as viewing $QH^\bullet_T(X)$ as a module for the action of the algebra of functions $\bC[p^{\pm1}, w]$ on $\bC^*_p \times \bC_w$ on, via \eqref{eq:0_diff_module} or alternatively \eqref{eq:glsm_spectral_curve}. Geometrically, the module forms a sheaf over $\bC^*_p \times \bC_w$  with holomorphic Lagrangian support $\cL$.

\subsubsection{Example: \texorpdfstring{$\mathbb{CP}^1$ $\sigma$-model.}{}}

For supersymmetric QED with $2$ chirals, from \eqref{eq:w_eff_sqed2}, the vacuum equations are:
\begin{equation}\label{eq:bethe}
1 = e^{\frac{\partial W_{\text{eff}}}{\partial \sigma}} = q^{-1} (\sigma+m)(\sigma-m),
\end{equation}
where $m \equiv iw/2$. This describes the quantum equivariant cohomology  $QH^\bullet_T(\mathbb{CP}^1)$. 

To obtain the automorphism $F(w)$ on a basis of $V$, $\{\ket{\mathbf{1}}, \sigma \ket{\mathbf{1}}\}$ generated by the twisted chiral ring basis $\{\mathbf{1},\sigma\}$, note
\begin{equation}\label{eq:0_diff_cp1}
    p =  e^{\frac{\partial W_{\text{eff}}}{\partial m}}  = \frac{\sigma+m}{\sigma-m} \quad \Rightarrow \quad p 
    \begin{pmatrix} 
    \mathbf{1} \\ 
    \sigma 
    \end{pmatrix} 
    = 
    F(m) 
    \begin{pmatrix} 
    \mathbf{1} \\ 
    \sigma 
    \end{pmatrix} 
\end{equation}
where
\begin{equation}\label{eq:sqed2_matrix_0_diff}
    F(m) = 
    \begin{pmatrix}
    1+2m^2 q^{-1} & 2m q^{-1} \\
    2m (1+ m^2q^{-1}) & 1+ 2m^2 q^{-1}
    \end{pmatrix}.
\end{equation}
The equality in the second equation in \eqref{eq:0_diff_cp1} should be considered up to the ring relation \eqref{eq:bethe}.

To derive the spectral curve, one can simply take the characteristic polynomial of the above, or alternatively solving for $\sigma$ in $p =  e^{\frac{\partial W_{\text{eff}}}{\partial m}} $ gives 
\begin{equation}
\sigma = \frac{m(p+1)}{p-1},
\end{equation}
and substituting into (\ref{eq:bethe}) we obtain
\begin{equation}\label{eq:spectralcurve}
\cL(m,p) \coloneqq p^2 - 2(1+2 m^2q^{-1} )p+ 1 = 0.
\end{equation}
It is easy to check the action of $p$ on $V$ (\textit{i.e.} $QH^\bullet_T(\mathbb{CP}^1)$) defined by \eqref{eq:0_diff_cp1} obeys \eqref{eq:spectralcurve}.

\subsection{\texorpdfstring{$\lambda \neq 0$}{}: branes, difference modules \& curve quantization}\label{sec:difference_modules}

We now consider the $\lambda \neq 0 $ case, corresponding to viewing the space of supersymmetric ground states as classes in $Q_\lambda$-cohomology. We first review how, in the work of Mochizuki \cite{mochizuki2017periodic}, the $0$-difference modules we discussed in the previous section are replaced by $2i \lambda $-difference modules. We then realize these physically via brane amplitudes and hemisphere partition functions.

Recall that $\lambda$ parametrises mini--complex structures on $S^1\times \bR$, which can be constructed by means of some $\lambda$-dependent mini--complex structures on a lift $\mathbb{R}_{t_1}\times \mathbb{C}_{\beta_1}$. It follows from~\eqref{eq:supercharge_dependencies} that the operators $\partial_{\bar{\beta_1}}$, $\partial_{t_1}$ descend to the space of supersymmetric ground states. By restricting to a constant value of $t_1$, we can then define the holomorphic vector bundle on $\bC_{\beta_1}$
\begin{equation}\label{eq:diff_module_bundle}
    \cE^{t_1} := (E|_{\{t_1\} \times\bC_{\beta_1}}, \partial_{E,\bar{\beta}_1} ).
\end{equation}
We can then consider the complex Bogomolny equation in these variables 
\begin{equation}\label{eq:complex_bogomolny_lambda}
    [\partial_{E,t_1}, \partial_{E,\bar{\beta}_1}] = 0.
\end{equation}
The difference operator is now defined as:
\begin{equation}
    \Phi_V^* = \Phi_1^* \circ F
\end{equation}
where $F: \cE^0 \rightarrow \cE^1$
is the endomorphism given by parallel transport with respect to $\partial_{E,t_1}$, and $\Phi_1^*$ is the pullback induced by the automorphism $\Phi_1 : \bC_{\beta_1} \rightarrow \bC_{\beta_1}$ given by $\Phi_1(\beta)= \beta_1+2i\lambda $, so that
\begin{equation}
\Phi_1^*(\mathcal{E}^1) \cong \mathcal{E}^0.
\end{equation}

Let $V_{\text{naïve}}$ be the $\bC(\beta_1)$-module of holomorphic sections of $\cE^0$
\begin{equation}
    V_{\text{na\"ive}}:= H^0 (\mathbb{C}_{\beta_1}, \cE^0)\otimes_{\mathbb{C}[\beta_1]}\mathbb{C}(\beta_1).
\end{equation}
To make this rigorous one needs to once again prescribe a certain behavior at infinity. Moreover, in the presence of singularities we must in general allow for meromorphic sections and prescribe a corresponding parabolic structure. We will not review these important technical details here, but simply assume that there is a well--defined $\mathbb{C}(w)$-module $V$, which replaces $V_{\text{naïve}}$.  The pair $(V, \Phi_V^*)$ of a $\mathbb{C}(\beta_1)$-module $V$ together with the automorphism $\Phi_1^*$ constitutes a $2i\lambda$-difference module. This means that if we have $f \in  \mathbb{C}(\beta_1)$ and $s \in V$, it follows from \eqref{eq:complex_bogomolny_lambda} that
\begin{equation}\label{eq:differencemodule}
\Phi_V^*(fs) = \Phi_1^*(f) \Phi_V^*(s).
\end{equation}

\subsubsection{Branes \& states}

In this section we relate ground states of the SQM along $\mathbb{R}$ of a cigar geometry to the $2i\lambda$-difference modules of Mochizuki. The cigar geometry is A--twisted in the bulk. We consider states generated by D--branes, which for our purposes are half--BPS boundary conditions for this configuration preserving $R_V$, and two supercharges $Q_{\lambda}$ and $\bar{Q}^{\dagger}_{\lambda}$, where $\lambda$ lies initially on the unit circle. Therefore such branes generate a harmonic, albeit not necessarily normalisable, representative of a state in $Q_{\lambda}$ cohomology, which we use to represent ground states. For $\lambda =1$, a linear combination of these supercharges is the B--type supercharge, and the corresponding D--branes are usually referred to as B--branes.

\begin{figure}
    \centering
    \tikzset{every picture/.style={line width=0.75pt}} 
\begin{tikzpicture}[x=0.75pt,y=0.75pt,yscale=-.5,xscale=.7]

\draw  [draw opacity=0][dash pattern={on 0.84pt off 2.51pt}] (450,160) .. controls (438.95,160) and (430,142.09) .. (430,120) .. controls (430,97.91) and (438.95,80) .. (450,80) -- (450,120) -- cycle ; \draw  [color={rgb, 255:red, 208; green, 2; blue, 27 }  ,draw opacity=1 ][dash pattern={on 0.84pt off 2.51pt}] (450,160) .. controls (438.95,160) and (430,142.09) .. (430,120) .. controls (430,97.91) and (438.95,80) .. (450,80) ;  
\draw  [draw opacity=0] (450,80) .. controls (450,80) and (450,80) .. (450,80) .. controls (450,80) and (450,80) .. (450,80) .. controls (461.05,80) and (470,97.91) .. (470,120) .. controls (470,142.09) and (461.05,160) .. (450,160) -- (450,120) -- cycle ; \draw  [color={rgb, 255:red, 208; green, 2; blue, 27 }  ,draw opacity=1 ] (450,80) .. controls (450,80) and (450,80) .. (450,80) .. controls (450,80) and (450,80) .. (450,80) .. controls (461.05,80) and (470,97.91) .. (470,120) .. controls (470,142.09) and (461.05,160) .. (450,160) ;  
\draw  [draw opacity=0] (236,160) .. controls (236,160) and (236,160) .. (236,160) .. controls (169.73,160) and (116,142.09) .. (116,120) .. controls (116,97.91) and (169.73,80) .. (236,80) -- (236,120) -- cycle ; \draw   (236,160) .. controls (236,160) and (236,160) .. (236,160) .. controls (169.73,160) and (116,142.09) .. (116,120) .. controls (116,97.91) and (169.73,80) .. (236,80) ;  
\draw    (370,80) -- (450,80) ;
\draw    (370,160) -- (450,160) ;
\draw  [line width=3]  (116,120.5) .. controls (116,120.22) and (116.22,120) .. (116.5,120) .. controls (116.78,120) and (117,120.22) .. (117,120.5) .. controls (117,120.78) and (116.78,121) .. (116.5,121) .. controls (116.22,121) and (116,120.78) .. (116,120.5) -- cycle ;
\draw  [dash pattern={on 4.5pt off 4.5pt}]  (236,80) -- (370,80) ;
\draw  [dash pattern={on 4.5pt off 4.5pt}]  (240,160) -- (370,160) ;

\draw (478,110) node [anchor=north west][inner sep=0.75pt]  [color={rgb, 255:red, 208; green, 2; blue, 27 }  ,opacity=1 ]  {$D$};
\draw (85,110) node [anchor=north west][inner sep=0.75pt]    {$\cO_a$};
\draw (271,172.4) node [anchor=north west][inner sep=0.75pt]    {$\braket{a|\textcolor[rgb]{0.82,0.01,0.11}{D}}$};
\end{tikzpicture}
    \caption{The brane amplitude given by the overlap between the state $\ket{D}$ generated by the brane, and $\bra{a}$ generated by the path integral with an insertion of a twisted chiral ring operator.}
    \label{fig:brane_amplitude}
\end{figure}
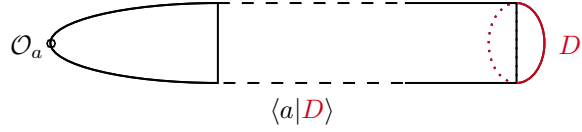
We denote by $\Pi[D]$ the projection of $\ket{D}$ onto the space of supersymmetric ground states. This can be done by taking inner products (via computing the path integral on the infinite cigar), yielding brane amplitudes. For example, we can consider the overlap
\vspace*{-2mm}
\begin{equation}\label{eq:overlap}
    \Pi[D,a] = \braket{a|D}
\end{equation}
where as before $\ket{a}$ is the ground state generated by the path integral for the topologically twisted theory on an indefinitely long cigar, with a (twisted) chiral ring operator $\cO_a$ labeled by $a$ inserted at the tip, as shown in Figure~\ref{fig:brane_amplitude}. 

As functions of $\lambda$, it is known\cite{Cecotti:1992rm,Gaiotto:2011tf} that the brane amplitudes can be analytically continued to the whole of $\bC\backslash\{0,\infty\}$. Further, it is a classic result in the context of $tt^*$ equations that $\Pi[D]$ are flat sections of the Lax connection:
\begin{equation}\label{eq:lax_equations}
\begin{aligned}
   L_{w,t} = D_{w} + \frac{i}{2\lambda}(D_{t}-i\phi),\qquad
   \bar{L}_{\bar{w},t} = \bar{D}_{\bar{w}} + \frac{i\lambda}{2}(D_{t}+i\phi),
\end{aligned}
\end{equation}
\textit{i.e.}  $L_{w,t} \Pi[D] = \bar{L}_{\bar{w},t}  \Pi[D] = 0$.
The key insight is then that one may rewrite the holomorphic covariant derivative and parallel transport operator of Mochizuki as:
\begin{equation}
    \partial_{E,t_1} = \frac{2i}{1+|\lambda|^2} \left(\bar{\lambda} \bar{L}_{\bar{w},t} - \lambda L_{w,t}\right), \qquad
    \partial_{E,\bar{\beta}_1} = \frac{1}{1+|\lambda|^2} \bar{L}_{\bar{w},t}.
\end{equation}

Although the expansion in the basis $\ket{a}$ is natural from the point of view of the $tt^*$,\cite{Ooguri:1996ck} in the following it will be useful to introduce a basis $\ket{a^\lambda}$ such that
\begin{itemize}
    \item $\ket{a^\lambda (t_1=0,\beta_1)}$ is a holomorphic section of $\cE^0$
    \item{$\lim_{\lambda \rightarrow 0 }\ket{a^\lambda} = \ket{a}$.}
\end{itemize}
Locally, we can always find a basis holomorphic of sections, and so a basis $\ket{a^\lambda}$ satisfying the first bullet point. We also know that as $\lambda \rightarrow 0$, the chiral ring basis is holomorphic. Therefore, without loss of generality, we can assume that the second bullet point holds. We can then expand
\begin{equation}
\Pi[D] := \sum_{a,b} \Pi[D,b^{\lambda}] \eta^{ab} \ket{a^\lambda}.
\end{equation}
where $\eta^{ab}$ is the inverse matrix of $\eta_{ab} = \braket{a^\lambda|b^\lambda}$.

Then, whenever they are well--defined, the flatness of the D--brane amplitudes under the Lax connection \eqref{eq:lax_equations} imply that restricting to $t_1= 0$, $\Pi[D]|_{t_1=0}$ is holomorphic in $\beta_1$, and thus $\Pi[D]|_{t_1=0}$ can in principle be identified with elements of the difference modules $V$.

Moreover (at least formally) the brane states are solutions to the parallel transport equations,  since
\begin{equation}
\partial_{E,t_1}  \Pi[D] = 0.
\end{equation}
We further assume that brane amplitudes are globally defined functions of $t$ (\textit{i.e.} are $t$-periodic).  This was shown to hold for the free chiral, and general LG theories. \cite{Cecotti:2013mba} More generally, any non--trivial behavior under shifting $t$ by an element of $\bZ^n$ arises due to 't Hooft anomalies involving $T$. For our GLSMs, the only such anomalies are mixed $T-R_A$ anomalies where $R_A$ is the axial $R$-symmetry. Since there is no non--trivial background for $R_A$, the shift in $t$ cannot produce any non--trivial phase in the partition function. Then, in terms of the original coordinates $(t,w,\bar{w})$, we can see from \eqref{eq:mochizukisecondcoords} that this becomes simply:
\begin{equation}\label{eq:brane_amplitude_transport}
\Phi_{V}^* : \Pi[D](t, w, \bar{w}) \rightarrow \Pi[D](t+ L, w, \bar{w}) = \Pi[D](t, w, \bar{w}).
\end{equation}

This means that the D--brane states are an invariant of the module action. Therefore, under our assumptions, computing a basis of brane amplitudes is equivalent to determining the module associated to the monopole representing the Berry connection. In fact, if we were able to find a basis of brane amplitudes for $V$, a general section $s$ of $\mathcal{E}^0$ could be expanded in terms of a $\mathbb{C}(\beta_1)$ linear combination of the brane amplitudes, then the action of the automorphism $\Phi_{V}^*$ on it is trivial to compute. The problem is of course that it is, in general, very difficult to compute the brane amplitudes explicitly since they are non--supersymmetric: they are $A$--twisted in the bulk yet preserve $Q_{\lambda},\bar{Q}_{\lambda}^{\dagger}$ at the boundary.

\subsubsection{Difference equations for brane amplitudes \& curve quantization}

In this section we derive from our previous considerations difference equations for brane amplitudes. We further demonstrate that in the $\lambda \rightarrow 0$ limit these difference equations recover the Cherkis--Kapustin spectral curve discussed in Section~\ref{sec:product_case}. 

Note that the automorphism $\Phi_{V}^*$ sends an element of $V$, \textit{i.e.} a holomorphic section of $\cE^0$ to another element of $V$. Therefore we can expand the action of  $\Phi_V^*$  on the basis $\{\ket{b^\lambda}\}$, the twisted chiral ring, by:
\begin{equation}
    \Phi_{V}^*\ket{a^{\lambda}(0,\beta_1)} = \sum_{b} {M^{b}_{a}}(\beta_1) \ket{b^\lambda(0,\beta_1)}
\end{equation}
where ${M_{a}^{b}}$ must be holomorphic in $\beta_1$. Using equation \eqref{eq:brane_amplitude_transport}, we have:
\begin{equation}
\begin{aligned}
    \sum_{a,b}\braket{b^{\lambda}|D}\eta^{ab}\ket{a^\lambda} 
    = \sum_{a,b,c} \Phi_{1}^*(\braket{b^\lambda|D}\eta^{ab}){M^{c}_{a}}\ket{c^\lambda} 
\end{aligned}
\end{equation}
where in the above, all objects are evaluated at $t_1=0$ and arbitrary $(\beta_1,\bar{\beta}_1)$. We conclude that if we regard $\braket{\cdot | D}$ as an $N$-vector with components $\braket{ a^\lambda | D}$ we can write:
\begin{equation}\label{eq:matrix_difference_equation}
    (\Phi_{1}^{*})^{-1} \braket{\cdot | D} =  G \, \braket{\cdot | D}.
\end{equation}
where $G$ is the holomorphic matrix $G_a^b(\beta_1) := (\Phi_{1}^*)^{-1}\left( \eta_{ad} M^{d}_{c}\right)\eta^{cb}$. This is a matrix difference equation, which to the best of our knowledge is novel. \footnote{To--date, it seems as though only \textit{differential} Picard--Fuchs equations arising from the $tt^*$ geometry associated to the K\"ahler (Fayet--Iliopoulos) parameter have been studied in 2d, see \textit{e.g.}\cite{Blok:1991bi, Lerche:1991wm, cadavid1991picard, Morrison:1991cd}.} It is a quantization of \eqref{eq:0_difference}, as we shall see in the next subsection.

For the moment, let us show that the difference equations provide a quantization of the Cherkis--Kapustin spectral curve. We make use of a particularly nice set of brane amplitudes, namely \textit{thimble branes} $D_\alpha$, whose boundary amplitudes give a fundamental basis of flat sections for the $tt^*$ Lax connection \cite{Cecotti:1992rm,Gaiotto:2011tf}. For LG models, they are Lagrangian submanifolds projecting to straight lines in the $W$-plane beginning at critical points $\alpha$ of $W$. For GLSMs which flow in the IR to NLSMs, they are the holomorphic submanifolds of $X$ corresponding to attracting submanifolds of fixed points (\textit{i.e.} vacua $\{\alpha\}$)  for the Morse flow generated by $w_2$. Such boundary conditions have been analyzed explicitly for massive $(2,2)$ theories \cite{Hori:2000ck, Gaiotto:2015aoa} and 3d $\mathcal{N}=4$ theories\cite{Bullimore:2016nji, Bullimore:2020jdq, Bullimore:2021rnr, Crew:2023tky}. 

Note that the difference equation \eqref{eq:matrix_difference_equation} holds for any B--brane $D$ is equivalent to it holding for the basis of thimble branes. This is because any brane amplitude can be written as a $\bZ$--linear combination of the $\{D_a\}$ amplitudes 
\begin{equation}\label{eq:thimbles_basis}
    \Pi[D] = \sum_{\alpha} n_{\alpha} \Pi[D_{\alpha}]
\end{equation}
where $n_{\alpha}$ are the framed BPS degeneracies \cite{Gaiotto:2011tf}.

A key fact we will make extensive use of is that the asymptotic behavior in $\lambda$ of the thimble brane amplitudes is known:
\begin{equation}\label{eq:thimble_brane_asymptotic}
    \braket{b^\lambda|D_\alpha} \sim e^{\frac{W_{\text{eff}}^{(\alpha)}}{\lambda}} \cO_b|_{\alpha},\quad\text{as } \lambda \rightarrow 0.
\end{equation}
In the above, the effective twisted superpotential is computed at an RG scale $\mu = \lambda$. Here, the effective twisted superpotential $W_{\text{eff}}^{(\alpha)}$ is evaluated at the vacuum solution in $\sigma$ which reduces to its value in the classical vacuum configuration in the limit where the (exponential of) the FI parameters goes to zero. This is discussed for supersymmetric QED in Section~\ref{subsubsec:sqed2_asymptotics}.

Let us now see in what sense this provides a quantization of the $\lambda=0$ spectral curve. If we denote $\braket{\mathbf{1}|D_{\alpha}}$ the thimble brane amplitude with the trivial (no) operator insertion, then from \eqref{eq:thimble_brane_asymptotic} we note that:
\begin{equation}
    \lim_{\lambda \rightarrow 0 }\frac{ (\Phi_{1}^{*})^{-1}\braket{a^{\lambda}|D_{\alpha}}}{\braket{\mathbf{1}|D_{\alpha}}} = e^{-2 i \frac{\partial W_{\text{eff}}^{(\alpha)}}{\partial w} } \cO_a|_{\alpha}.
\end{equation}
In the above, we have traded $(\Phi_{1}^{*})^{-1}$ for a shift $w \rightarrow w-2i\lambda  $, which is valid in the $\lambda \rightarrow 0$ limit due to the definitions \eqref{eq:mochizukisecondcoords}. Here $\cO_a|_{\alpha}$ denotes the evaluation of the operator $\cO_a$ in the vacuum $\alpha$. Note that $\exp{-2 i \partial_{w} W_{\text{eff}}^{(\alpha)}} $ for $\alpha =1,\ldots,N$ are precisely the solutions for $p$ in the spectral curve. Using \eqref{eq:matrix_difference_equation}, we conclude that:
\begin{equation}
    \lim_{\lambda \rightarrow 0 } \frac{\cL(w, (\Phi_{1}^{*})^{-1}) \braket{\cdot |D_{\alpha}}}{\braket{\mathbf{1}|D_{\alpha}}} =    \lim_{\lambda \rightarrow 0 } \frac{\cL(w, \{G\}) \braket{\cdot |D_{\alpha}}}{\braket{\mathbf{1}|D_{\alpha}}} = 0.
\end{equation}
Since this holds for the basis of thimble amplitudes $\braket{\cdot|D_{\alpha}}$, we conclude that:
\begin{equation}
   \lim_{\lambda \rightarrow 0}  \cL (w, G) =0.
\end{equation}
We therefore recover the Cherkis--Kapustin spectral curve.

\subsection{Difference equations for hemisphere partition functions}\label{sec:conf_vortex}

We have remarked above that D--brane amplitudes are difficult to compute in general. In the conformal limit these are however expected to degenerate into exactly calculable hemisphere partition functions. \cite{Cecotti:2013mba} The limit corresponds to taking 
\begin{equation}\label{eq:conf-lim}
   \mathrm{lim}_c :\quad \lambda \rightarrow 0,\quad L \rightarrow 0,\quad \lambda/L = \ep,
\end{equation}
where $\ep$ is an arbitrary constant and $L$ is the radius of the circle of the cylinder on which our system is quantized. 

We re--introduce $L$ in this section, and define the complex mass $m$ and normalized holonomy $t'$ (with period $1$) via $w = -2iL^2 m$ and $t = L t'$, so that in:
\begin{gather}
    \beta_1 = -2 i L^2 m + 2 i \lambda L t' + 2 i \lambda^2 L^2 \bar{m}, \qquad 
    t_1 = L t' + \text{Im}(  2i \lambda L^2 \bar{m})
\end{gather}
the dimensions of the summands are consistent. Thus:
\begin{equation}
    \mathrm{lim}_c\, \Pi[D,a](t_1=0,\beta_1,\bar{\beta}_1) = \mathcal{Z}_D[\cO_a, m-\epsilon t'].
\end{equation}
Here $\mathcal{Z}_D[\cO_a]$ denotes the hemisphere partition function with boundary condition $D$ on $S^1 = \partial HS^2$, and a twisted chiral ring operator $\cO_a$ inserted at the tip of the hemisphere. The radius of the hemisphere is given by $\epsilon^{-1}$. In our conventions, $m-\epsilon t'$ appears in place of the usual complex mass deformation $m$ in the hemisphere partition function, \cite{Honda:2013uca} as we shall see in our examples. We replace this combination simply by $m$ in the following. The hemisphere partition functions \textit{are} BPS objects that can be computed via localization~\cite{Hori:2013ika, Honda:2013uca,Sugishita:2013jca}, and are holomorphic in $m$. They are also equivalent to the vortex partition functions on $\bR^2_{\ep}$ computed in an Omega background \cite{Fujimori:2015zaa}, with the Omega deformation parameter $\ep$.

Let us briefly recap why the D--brane amplitudes are expected to degenerate into the hemisphere partition functions in this limit. In the conformal limit:
\begin{equation}
     L_{w,t} \rightarrow D_w + \frac{i}{2 \epsilon}(D_t-i\phi),\qquad \bar{L}_{\bar{w},t} \rightarrow \partial_{\bar{w}}.
\end{equation}
In the second limit we have used the fact we are working in topological gauge $(\bar{A}_{\bar{w}})_a{}^{b}=0$. This is consistent with the holomorphy of hemisphere partition functions in the complex mass. For LG models, the solutions to such equations are given by period integrals \cite{Hori:2000ck}, which have been shown to equal the results of localization for their mirror dual GLSMs \cite{Fujimori:2012ab}. Later, we also compute the hemisphere (vortex) partition functions for some examples, and verify they satisfy the conformal limit of the difference equations noted above for D--brane amplitudes, giving further support for this claim. See also \cite{Cecotti:2013mba} for the the explicit example of the free chiral. 

\subsubsection{Difference equations}\label{sec:finite_difference_HPS}

Let us now write out explicitly the difference equations for hemisphere or vortex partition functions. They are simply the conformal limit of \eqref{eq:matrix_difference_equation}. Let us substitute $m-\ep t' \rightarrow m$ as above. Denoting
\vspace{-1mm}
\begin{equation}\label{eq:diff_operators}
    \hat{p}= e^{\epsilon \partial_{m}},\quad\hat{m} = \times m
\end{equation}
then:
\vspace{-2mm}
\begin{equation}
[\hat{p},\hat{m} ]= \epsilon  \hat{p} \quad\Rightarrow\quad \hat{p}: m \mapsto m+\epsilon.
\end{equation}
Thus $\hat{p}$ is a difference operator for $m$, and coincides with the operator $(\Phi_{1}^{*})^{-1}:\beta_1 \mapsto \beta_1-2i\lambda L$ in the conformal limit. Therefore we obtain difference equations
\begin{equation}\label{eq:matrix_difference_vortex}
    \hat{p}\, \mathcal{Z}_D[\cO_a, m ] = \mathcal{Z}_D[\cO_a, m + \ep] =  \widetilde{G}_{ab}(m,\ep) \mathcal{Z}_D[\cO_b, m ].
\end{equation}
Here $\widetilde{G}= \mathrm{lim}_c\,G$, where $G$ is the matrix appearing in the difference equations for brane amplitudes \eqref{eq:matrix_difference_equation}. 

This yields an $\ep$-deformation of the $0$-difference modules $\eqref{eq:0_difference}$, and therefore exhibits $QH_T(X)$ as a module for the quantized algebra of functions $\bC[\hat{p}^{\pm1}, \hat{w}]$. In particular:
\begin{equation}\label{eq:limit_matrix}
    \lim_{\ep \rightarrow 0 }\widetilde{G}(m,\ep) = F(m)
\end{equation}
where $F$ is the automorphism appearing in the $0$-difference module \eqref{eq:0_difference}. To the best of our knowledge, this is a new result on difference relations satisfied by these partition functions.

As for D--branes, that \eqref{eq:matrix_difference_vortex} holds for any B--brane is equivalent to it holding for each of the thimble branes. The $\epsilon \rightarrow 0$ behavior of the hemisphere partition functions equipped with the thimble brane boundary conditions $\{D_{\alpha}\}$ can be derived from the asymptotic behavior of the thimble brane amplitudes \eqref{eq:thimble_brane_asymptotic}:
\begin{equation}
    \cZ_{D_{\alpha}}[\cO_b,m] \sim e^{\frac{W_{\text{eff}}^{(\al)}}{\ep}} \cO_b, \quad \text{as } \ep \rightarrow 0.
\end{equation}
This is consistent with the limit for thimble brane amplitudes as upon reintroducing the circle length $L$ the superpotential is rescaled $W \rightarrow L W$. The same arguments as for brane amplitudes let us conclude that 
\begin{equation}\label{eq:vortex_spectral_curve}
    \lim_{\ep \rightarrow 0} \cL( m, \widetilde{G}) = 0,
\end{equation}
and by Cayley--Hamilton that $\widetilde{G}$ has eigenvalues $\exp \partial_{m} W_{\text{eff}}^{(\alpha)}$. Thus we obtain a quantization of the Cherkis--Kapustin spectral curve. 

To show~\eqref{eq:limit_matrix} directly, we can import Coulomb branch localization formulae \cite{Honda:2013uca, Hori:2013ika, Sugishita:2013jca} that express $\mathcal{Z}_D[\cO_a, m ]$ as a contour integral over the Coulomb branch scalars $\sigma$. $\cO_a$ is represented by a polynomial in $\sigma$. The integrand scales as $\ep \rightarrow 0$ as $e^{W_{\text{eff}}[\sigma,m]/\ep}$, and so in the integral and the limit,  $\hat{p}$ acts precisely as multiplication by $e^{\partial_{m} W_{\text{eff}}[\sigma,m]}$, recovering its action in the $0$-difference module case, as described in section \ref{sec:physical_examples}.

Beautifully, as hemisphere and vortex partition functions are calculable via localization \cite{Honda:2013uca,Hori:2013ika,Sugishita:2013jca}, this gives a recipe, arising from 2d GLSMs, to construct solutions (involving hypergeometric functions) to difference equations arising as quantized spectral curves (in turn corresponding to quantum equivariant cohomologies of K\"ahler varieties). Note that hemisphere partition functions can be interpreted as equivariant Gromov--Witten invariants of the Higgs branches \cite{Bonelli:2013mma}. 

\subsubsection{Example: \texorpdfstring{$\mathbb{C}\mathbb{P}^1$}{}}

\begin{figure}   \centering

  
\tikzset {_4em7x9s1j/.code = {\pgfsetadditionalshadetransform{ \pgftransformshift{\pgfpoint{89.1 bp } { -128.7 bp }  }  \pgftransformscale{1.32 }  }}}
\pgfdeclareradialshading{_284suey81}{\pgfpoint{-72bp}{104bp}}{rgb(0bp)=(1,1,1);
rgb(0bp)=(1,1,1);
rgb(25bp)=(0.82,0.01,0.11);
rgb(400bp)=(0.82,0.01,0.11)}
\tikzset{_ar3t0txhx/.code = {\pgfsetadditionalshadetransform{\pgftransformshift{\pgfpoint{89.1 bp } { -128.7 bp }  }  \pgftransformscale{1.32 } }}}
\pgfdeclareradialshading{_jhb5n2thn} { \pgfpoint{-72bp} {104bp}} {color(0bp)=(transparent!0);
color(0bp)=(transparent!0);
color(25bp)=(transparent!40);
color(400bp)=(transparent!40)} 
\pgfdeclarefading{_evzegvolg}{\tikz \fill[shading=_jhb5n2thn,_ar3t0txhx] (0,0) rectangle (50bp,50bp); } 

  
\tikzset {_vgnoi74zt/.code = {\pgfsetadditionalshadetransform{ \pgftransformshift{\pgfpoint{89.1 bp } { -128.7 bp }  }  \pgftransformscale{1.32 }  }}}
\pgfdeclareradialshading{_ukzk2nnf3}{\pgfpoint{-72bp}{104bp}}{rgb(0bp)=(1,1,1);
rgb(1.607142857142857bp)=(1,1,1);
rgb(25bp)=(0.61,0.61,0.61);
rgb(400bp)=(0.61,0.61,0.61)}
\tikzset{_61sun0l58/.code = {\pgfsetadditionalshadetransform{\pgftransformshift{\pgfpoint{89.1 bp } { -128.7 bp }  }  \pgftransformscale{1.32 } }}}
\pgfdeclareradialshading{_53cfrpsln} { \pgfpoint{-72bp} {104bp}} {color(0bp)=(transparent!0);
color(1.607142857142857bp)=(transparent!0);
color(25bp)=(transparent!44.99999999999999);
color(400bp)=(transparent!44.99999999999999)} 
\pgfdeclarefading{_gnx4dp8jq}{\tikz \fill[shading=_53cfrpsln,_61sun0l58] (0,0) rectangle (50bp,50bp); } 
\tikzset{every picture/.style={line width=0.75pt}} 

\begin{tikzpicture}[x=0.44pt,y=0.44pt,yscale=-1,xscale=1]

\path  [shading=_284suey81,_4em7x9s1j,path fading= _evzegvolg ,fading transform={xshift=2}] (101.5,130.75) .. controls (101.5,92.5) and (132.5,61.5) .. (170.75,61.5) .. controls (209,61.5) and (240,92.5) .. (240,130.75) .. controls (240,169) and (209,200) .. (170.75,200) .. controls (132.5,200) and (101.5,169) .. (101.5,130.75) -- cycle ; 
 \draw  [color={rgb, 255:red, 0; green, 0; blue, 0 }  ,draw opacity=1 ] (101.5,130.75) .. controls (101.5,92.5) and (132.5,61.5) .. (170.75,61.5) .. controls (209,61.5) and (240,92.5) .. (240,130.75) .. controls (240,169) and (209,200) .. (170.75,200) .. controls (132.5,200) and (101.5,169) .. (101.5,130.75) -- cycle ; 

\path  [shading=_ukzk2nnf3,_vgnoi74zt,path fading= _gnx4dp8jq ,fading transform={xshift=2}] (370,130.75) .. controls (370,92.5) and (401,61.5) .. (439.25,61.5) .. controls (477.5,61.5) and (508.5,92.5) .. (508.5,130.75) .. controls (508.5,169) and (477.5,200) .. (439.25,200) .. controls (401,200) and (370,169) .. (370,130.75) -- cycle ; 
 \draw   (370,130.75) .. controls (370,92.5) and (401,61.5) .. (439.25,61.5) .. controls (477.5,61.5) and (508.5,92.5) .. (508.5,130.75) .. controls (508.5,169) and (477.5,200) .. (439.25,200) .. controls (401,200) and (370,169) .. (370,130.75) -- cycle ; 

\draw  [color={rgb, 255:red, 0; green, 0; blue, 0 }  ,draw opacity=1 ][fill={rgb, 255:red, 208; green, 2; blue, 27 }  ,fill opacity=1 ] (96.5,130.75) .. controls (96.5,127.99) and (98.74,125.75) .. (101.5,125.75) .. controls (104.26,125.75) and (106.5,127.99) .. (106.5,130.75) .. controls (106.5,133.51) and (104.26,135.75) .. (101.5,135.75) .. controls (98.74,135.75) and (96.5,133.51) .. (96.5,130.75) -- cycle ;
\draw  [fill={rgb, 255:red, 0; green, 0; blue, 0 }  ,fill opacity=1 ] (235,130.75) .. controls (235,127.99) and (237.24,125.75) .. (240,125.75) .. controls (242.76,125.75) and (245,127.99) .. (245,130.75) .. controls (245,133.51) and (242.76,135.75) .. (240,135.75) .. controls (237.24,135.75) and (235,133.51) .. (235,130.75) -- cycle ;
\draw  [fill={rgb, 255:red, 0; green, 0; blue, 0 }  ,fill opacity=1 ] (365,130.75) .. controls (365,127.99) and (367.24,125.75) .. (370,125.75) .. controls (372.76,125.75) and (375,127.99) .. (375,130.75) .. controls (375,133.51) and (372.76,135.75) .. (370,135.75) .. controls (367.24,135.75) and (365,133.51) .. (365,130.75) -- cycle ;
\draw  [fill={rgb, 255:red, 208; green, 2; blue, 27 }  ,fill opacity=1 ] (503.5,130.75) .. controls (503.5,127.99) and (505.74,125.75) .. (508.5,125.75) .. controls (511.26,125.75) and (513.5,127.99) .. (513.5,130.75) .. controls (513.5,133.51) and (511.26,135.75) .. (508.5,135.75) .. controls (505.74,135.75) and (503.5,133.51) .. (503.5,130.75) -- cycle ;
\draw  [draw opacity=0][dash pattern={on 0.84pt off 2.51pt}] (170.75,61.5) .. controls (187.32,61.5) and (200.75,92.5) .. (200.75,130.75) .. controls (200.75,169) and (187.32,200) .. (170.75,200) -- (170.75,130.75) -- cycle ; \draw  [dash pattern={on 0.84pt off 2.51pt}] (170.75,61.5) .. controls (187.32,61.5) and (200.75,92.5) .. (200.75,130.75) .. controls (200.75,169) and (187.32,200) .. (170.75,200) ;  
\draw    (200,40) -- (143,40) ;
\draw [shift={(140,40)}, rotate = 360] [fill={rgb, 255:red, 0; green, 0; blue, 0 }  ][line width=0.08]  [draw opacity=0] (8.93,-4.29) -- (0,0) -- (8.93,4.29) -- cycle    ;
\draw    (470,40) -- (413,40) ;
\draw [shift={(410,40)}, rotate = 360] [fill={rgb, 255:red, 0; green, 0; blue, 0 }  ][line width=0.08]  [draw opacity=0] (8.93,-4.29) -- (0,0) -- (8.93,4.29) -- cycle    ;
\draw  [draw opacity=0][dash pattern={on 0.84pt off 2.51pt}] (440,61.5) .. controls (440,61.5) and (440,61.5) .. (440,61.5) .. controls (456.57,61.5) and (470,92.5) .. (470,130.75) .. controls (470,169) and (456.57,200) .. (440,200) -- (440,130.75) -- cycle ; \draw  [dash pattern={on 0.84pt off 2.51pt}] (440,61.5) .. controls (440,61.5) and (440,61.5) .. (440,61.5) .. controls (456.57,61.5) and (470,92.5) .. (470,130.75) .. controls (470,169) and (456.57,200) .. (440,200) ;  

\draw (70,122.4) node [anchor=north west][inner sep=0.75pt]  [color={rgb, 255:red, 208; green, 2; blue, 27 }  ,opacity=1 ]  {$v_{1}$};
\draw (249,122.4) node [anchor=north west][inner sep=0.75pt]    {$v_{2}$};
\draw (338,122.4) node [anchor=north west][inner sep=0.75pt]    {$v_{1}$};
\draw (518,122.4) node [anchor=north west][inner sep=0.75pt]  [color={rgb, 255:red, 208; green, 2; blue, 27 }  ,opacity=1 ]  {$v_{2}$};
\draw (100,219.4) node [anchor=north west][inner sep=0.75pt]    {$D_{1} :\ \mathbb{CP}^{1} -\{v_{2}\}$};
\draw (395,222.4) node [anchor=north west][inner sep=0.75pt]    {$D_{2} :\{v_{2}\}$};

\end{tikzpicture}
    \caption{The support of the thimble boundary conditions for vacua $v_1$ and $v_2$ for supersymmetric QED with two chirals, \textit{i.e.} the $\mathbb{CP}^1$ sigma model. The arrow indicates the direction of Morse flow.}
    \label{fig:sqed2_thimbles}
\end{figure}
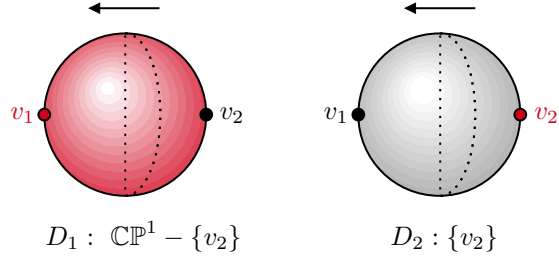

We return to our example of supersymmetric QED with two chirals, which flows to a non--linear sigma model to $\mathbb{CP}^1$ in the IR. For the sake of brevity, we work in a fixed chamber $\text{Re}(m) > 0$. We will denote the vacua $v_1$ and $v_2$, for which the thimble branes should be supported in the NLSM on (see Figure \ref{fig:sqed2_thimbles}):
\begin{equation}
    D_1\,:\,\mathbb{CP}^1 - \{v_2\}, \quad D_2\,:\,\{v_2\}.
\end{equation}
As before the two chirals $\Phi_1$ and $\Phi_2$ have charges $(+1,\pm 1)$ under $G\times T$ respectively. Only $\Phi_1$ obtains a VEV in vacuum $v_1$, and $\Phi_2$ in $v_2$. In the UV, the thimble branes are engineered by assigning the following boundary conditions to the chirals
\begin{equation}
    D_1\,:\, \Phi_1,\Phi_2 \text{ Neumann}, \quad 
    D_2\,:\, \Phi_1 \text{ Dirichlet},\, \Phi_2 \text{ Neumann}.
\end{equation}
In the opposite chamber, boundary conditions are given by exchanging $1\leftrightarrow 2$ in the above. The twisted chiral ring of supersymmetric QED is generated by $\{\mathbf{1},\sigma\}$ and is subject to the relation \eqref{eq:bethe}.

We now proceed to compute the hemisphere partition functions equipped with these boundary conditions, and demonstrate the quantization \eqref{eq:matrix_difference_vortex}~--~\eqref{eq:vortex_spectral_curve} of the spectral curve \eqref{eq:spectralcurve}. The partition functions are given by the contour integrals\cite{Honda:2013uca}
\begin{equation}\label{eq:sqed2_hps}
\begin{aligned}
    \mathcal{Z}_{D_1}[\cO_a] &= 
    \oint_{\mathcal{C}_1} \frac{d\sigma}{2\pi i \ep}
    e^{-\frac{2\pi i \sigma \tau}{\ep}} \,\,\Gamma\left[\frac{\sigma+m}{\ep}\right]
    \Gamma\left[\frac{\sigma-m}{\ep}\right] \cO_a,\\
    \mathcal{Z}_{D_2}[\cO_a] &= 
    \oint_{\mathcal{C}_2} 
    \frac{d\sigma}{2\pi i \ep}
    e^{-\frac{2\pi i \sigma \tau}{\ep}} \,\,
    \frac{(-2\pi i)e^{ \frac{\pi i (\sigma+m)}{\ep}}}{\Gamma\left[1-\frac{\sigma+m}{\ep}\right]}
    \Gamma\left[\frac{\sigma-m}{\ep}\right] \cO_a. 
\end{aligned}
\end{equation}
where $\mathcal{C}_1$ encloses the poles at $\sigma = -\ep k - m$, $k \in \mathbb{N}_0$, and $\mathcal{C}_2$ encloses the poles at $\sigma = -\ep k + m$.  Further $\cO_0= \mathbf{1}$, $\cO_1 = \sigma$, and $\tau = \tau(\ep)= \tau_0 + \frac{2}{2\pi i }\log(\Lambda_0/\ep)$ is the renormalized FI parameter at energy $\mu=\ep$. Note that $\cZ_{D_1}$ coincides with the vortex partition function computed in the Omega background $\bR_{\ep}^2$ \cite{Fujimori:2015zaa}.

It is easy to compute the contour integrals for $\mathcal{Z}_{D_1}[\cO_a]$
\begin{equation}
\begin{gathered}
    \mathcal{Z}_{D_1}[\mathbf{1}] = e^{\frac{2\pi i m \tau}{\epsilon}}\Gamma\left[-\frac{2m}{\epsilon}\right] {}_{0}F_1\left[1+\frac{2m}{\epsilon} ; e^{2\pi i \tau}\right], \\ 
    \mathcal{Z}_{D_1}[\sigma] = -m \mathcal{Z}_{D_1}[\mathbf{1}]  + \ep \, e^{2\pi i \tau} e^{\frac{2\pi i m \tau}{\ep}} \Gamma\left[-1-\frac{2m}{\ep}\right]{}_0F_1\left[2+\frac{2m}{\ep}; e^{2\pi i \tau}\right].\\ 
\end{gathered}
\end{equation}
where ${}_0F_1$ is the standard generalized hypergeometric function.\footnote{ For the second vacuum, one can use the Euler reflection formula to invert the first Gamma function in the contour integral to find that
\begin{equation}
    \mathcal{Z}_{D_2}[\cO_a] 
    = \left(1-e^{\frac{4\pi i m}{\ep}}\right) \mathcal{Z}_{D_1}[\cO_a]|_{m\rightarrow -m} 
\end{equation}
}

Using the standard identity
\begin{equation}\label{eq:hypergeom_identity}
    {}_{0}F_1(b,z) = {}_{0}F_1(b+1,z) + \frac{z}{b(b+1)} {}_{0}F_1(b+2,z),
\end{equation}
we have
\begin{equation}\label{eq:matrix_diff_sqed2}
    \hat{p} 
    \begin{pmatrix} 
    \mathcal{Z}_{D_\alpha}[\mathbf{1}] \\ 
    \mathcal{Z}_{D_\alpha}[\sigma]
    \end{pmatrix} 
    = 
    \widetilde{G}(m,\ep) 
    \begin{pmatrix} 
    \mathcal{Z}_{D_\alpha}[\mathbf{1}] \\ 
    \mathcal{Z}_{D_\alpha}[\sigma]
    \end{pmatrix}
\end{equation}
where
\begin{equation}\label{eq:sqed2_matrix}
    \widetilde{G}(m,\ep) = 
    \begin{pmatrix}
    1+m(2m+\ep)q^{-1} & (2m+\ep)q^{-1} \\
    (2m+\ep)(1+ m(m+\ep)q^{-1}) & 1+ (m+\ep)(2m+\ep)q^{-1}
    \end{pmatrix},
\end{equation}
is the \textit{same} matrix for both vacua $\alpha$.  We have defined as before $q= \ep^2 e^{2 \pi i \tau}$. Notice that $\widetilde{G}(m,0) = F(m)$ where $F(m)$ is the corresponding automorphism in the $0$-difference case \eqref{eq:sqed2_matrix_0_diff}. Thus \eqref{eq:matrix_diff_sqed2} provides a quantization of the corresponding action on $QH_T(\mathbb{CP}^1)$.  Notice also that
\begin{equation}
   \lim_{\ep \rightarrow 0} \cL( m, \widetilde{G}(m,\ep)) 
   = 0
\end{equation}
where $\cL$ defines the monopole spectral curve \eqref{eq:spectralcurve}.

\section{Further Research}

We offer here some directions for future research.  It would be interesting to investigate what class of generalized periodic monopoles of GCK--type can be engineered as Berry connections of 2d GLSMs, and  interpreted in light of our results. It would also be interesting to explore the action of T--duality, and the gauging of global symmetries (corresponding to the Nahm transform on the Berry connection).

All of the structures should lift to counterparts for theories, for which Berry connections have been studied in depth.\cite{Cecotti:2013mba, Bullimore:2021rnr, Dedushenko:2021mds, Dedushenko:2022pem}. We expect that our difference equations for hemisphere partition functions are dimensional reductions of the qKZ equations obeyed by 3d hemisphere partition functions, which correspond to vertex functions in enumerative geometry\cite{Okounkov:2015spn, Crew:2023tky, Dedushenko:2023qjq}. In~\cite{Aganagic:2017smx} a physical origin of these difference equations is provided via compactifications of little string theory. Our results give a purely two (or three) dimensional construction, applying to theories which are not obtainable via such compactifications. The difference equations we would obtain in 3d should underlie the line operator identities obeyed by holomorphic blocks \cite{Beem:2012mb}.

Finally, as mentioned in the introduction, the results presented here are the physical manifestation of one (de Rham) side of a Riemann--Hilbert correspondence. A physical interpretation of the other (Betti) side was given in \cite{ad}. It would be interesting to physically study the correspondence further, and investigate the 3d lift, the generalized cohomology theories appearing there being $QK_T(X)$ and $\mathrm{Ell}_T(X)$.

\bibliographystyle{ws-ijmpa}
\bibliography{berryconnections}

\end{document}

%% file: preamble.tex
\usepackage{amsfonts}
\usepackage{amscd}
\usepackage{amssymb}
\usepackage{amsmath,bbm}
\usepackage{graphicx}
\usepackage{epsfig}
\usepackage{latexsym}
\usepackage{mathtools}
\usepackage{hyperref}
\usepackage{tikz-cd}
\usepackage[vcentermath]{youngtab}
    
\def \be  {\begin{equation}}
\def \ee  {\end{equation}}
\def \bea {\begin{equation}\begin{aligned}}
\def \eea {\end{aligned}\end{equation}}
\def \ba  {\begin{eqnarray}}
\def \ea  {\end{eqnarray}}
\def \bb  {\begin{thebibliography}}
\def \eb  {\end{thebibliography}}
\def \lab #1 {\label{#1}}
 \def\Im{{\rm Im}}

\newcommand\ep{\epsilon}

\newcommand\cE{\mathcal{E}}

\newcommand\cL{\mathcal{L}}

\newcommand\cN{\mathcal{N}}
\newcommand\cO{\mathcal{O}}

\newcommand\cZ{\mathcal{Z}}
\newcommand\al{\alpha}

\newcommand\bC{\mathbb{C }}

\newcommand\bR{\mathbb{R}}

\newcommand\bZ{\mathbb{Z}}

\definecolor{cardinal}{rgb}{0.6,0,0}
\definecolor{darkgreen}{rgb}{0,0.5,0}
\definecolor{golden}{rgb}{0.92, 0.7, 0}
\definecolor{midnight}{rgb}{0, 0, 0.5}
\definecolor{darkblue}{rgb}{0.2, 0, 0.8}